\documentclass{aastex63}

\def\a{\"a}
\def\o{\"o}
\def\u{\"u}
\def\A{\"A}
\def\O{\"O}
\def\U{\"U}
\def\etal{et al.~}

\def\ang{\AA}

\def\gapprox{\lower.4ex\hbox{$\;\buildrel >\over{\scriptstyle\sim}\;$}}
\def\lapprox{\lower.4ex\hbox{$\;\buildrel <\over{\scriptstyle\sim}\;$}}
\def\ref#1{\par\noindent\hangindent1cm {#1}}

\begin{document}

\title{Correlation of the sunspot number and the waiting time 
distribution of solar flares, coronal mass ejections, and 
solar wind switchback events observed with the Parker Solar Probe}

\correspondingauthor{Markus J. Aschwanden}

\author{Markus J. Aschwanden}
\email{aschwanden@lmsal.com}

\affiliation{Solar and Stellar Astrophysics Laboratory (LMSAL),
 Palo Alto, CA 94304, USA}

\author{Thierry Dudok de Wit}
\email{ddwit@cnrs-orleans.fr} 

\affiliation{Laboratory de Physique et Chimie de l'Environnement 
et de l'Espace, LPC2E, CNRS/CNES/University of Orl\'eans, 
3A Av. de la Recherche Scientifique, 45071 Orl\'eans cedex 2, France}

\begin{abstract}
Waiting time distributions of solar flares and {\sl coronal 
mass ejections (CMEs)} exhibit power law-like distribution 
functions with slopes in the range of 
$\alpha_{\tau} \approx 1.4-3.2$, as observed
in annual data sets during 4 solar cycles (1974-2012). 
We find a close correlation 
between the waiting time power law slope  
$\alpha_\tau$ and the {\sl sunspot number (SN)},
i.e., $\alpha_\tau$ = 1.38 + 0.01 $\times$ SN.
The waiting time distribution can be fitted
with a Pareto-type function of the form  
$N(\tau) = N_0$ $(\tau_0 + \tau)^{-\alpha_{\tau}}$, 
where the offset $\tau_0$ depends on the 
instrumental sensitivity, the detection threshold of 
events, and pulse pile-up effects. 
The time-dependent power law slope $\alpha_{\tau}(t)$ 
of waiting time distributions depends only 
on the global solar magnetic flux (quantified by 
the sunspot number) or flaring rate, independent of 
other physical parameters of {\sl self-organized 
criticality (SOC)} or {\sl magneto-hydrodynamic (MHD)}
turbulence models. 
Power law slopes of $\alpha_{\tau}\approx 1.2-1.6$ 
were also found in solar wind switchback events, 
as observed with the {\sl Parker Solar Probe (PSP)}. 
We conclude that the annual variability of switchback 
events in the heliospheric solar wind is modulated by 
flare and CME rates originating in the photosphere
and lower corona. 
\end{abstract} 


\section{	Introduction			}

Waiting times, also called {\sl elapsed times}, 
{\sl inter-occurrence times}, {\sl inter-burst times},
or {\sl laminar times}, are defined by the time interval
between two subsequent events, i.e., $\tau_i=
(t_{i+1}-t_{i})$, measured from a time series
$t_i=1,...,n_t$ of events. Simple examples are:
(i) periodic processes (where the waiting time 
is constant and is equal to the time period);
(ii) random, Poissonian, or Markov point processes 
(where the distribution of waiting times follows 
an exponential function); (iii) exponentially growing
avalanche processes (where the waiting time distribution
matches a scale-free power law-like function, as
it is common in {\sl self-organized criticality (SOC)} processes),
(iv) {\sl magneto-hydrodynamic (MHD)} turbulence processes 
(where power spectra
can be represented by piece-wise power law functions 
of waiting time distributions, or (v) sympathetic flaring, 
which is an effect that is not consistent with independent
flaring events (Wheatland et al.~1998). 
Hence, the study of
waiting time distributions, applied to solar flares here,
can be a powerful tool to identify and disentangle the
relevant physical processes, in particular in connection
with physical scaling laws (Aschwanden 2020). 

In solar physics, the waiting distribution functions of
solar flares has been found to be dominantly power 
law-like (Boffeta et al.~1999; Lepreti et al.~2001; 
Grigolini et al.~2002; Wheatland 2000a, 2003; 
Aschwanden and McTiernan 2010), unless the sample of
waiting times covers a too small range or is 
incompletely sampled otherwise, due to selection effects
(e.g., in Pearce et al.~1993; Crosby 1996), as demonstrated in
comparison with larger and more complete data sets
(Aschwanden and McTiernan 2010). The observationally
established result of power law functions in the waiting 
time distribution of solar flares rules out a stationary 
Poisson process and requires an alternative explanation in 
terms of non-stationary Poisson processes (Wheatland 2000a;
2003), self-organized criticality (SOC) models (Aschwanden
and McTiernan 2010; Aschwanden and Freeland 2012; Aschwanden et al.~2014),
or MHD turbulence (Boffetta et al.~1999; Lepreti et al.~2001;
Grigolini et al.~2002). It was argued that SOC avalanches 
occur statistically independently (Bak et al.~1987, 1988), 
and thus would predict 
an exponential waiting time distribution (Boffetta et al.~1999). 
However, inclusion of a driver with time-dependent variations,
such as the solar cycle variability (Aschwanden 2011b), leads to a
non-stationary Poisson process with power law behavior
(Wheatland 2000a, 2003). In this Paper we study the
time variability of the power law slope of waiting times
in more detail and find a strong correlation between the 
level of solar activity in terms of the sunspot number SN(t),
the annual flaring rate $\lambda(t)$, and the power law
slope $\alpha_\tau(t)$ of the waiting time distribution,
which can be explained by their common magnetic drivers. 
Obviously, the variability of solar activity represents
the most dominant physical process that modulates 
the value of the power law slope $\alpha_\tau(t)$ in solar flare
waiting time distributions, an argument that has not 
received much attention (except in 
Wheatland and Litvinenko 2002; Wheatland 2003),
but it puts previous modeling of waiting time
distributions into a new light.

New results emerge also from the variability of the solar wind, 
especially from the renewed interest in switchback events, as 
observed by the {\sl Parker Solar Probe (PSP)} mission. Switchbacks are 
sudden deflections of the magnetic field that have been found to 
be ubiquitous in the inner heliosphere. These events are likely 
to play an important role in structuring the young solar wind 
(Mozer et al.~2002; Tenerani et al.~2020; Horbury et al.~2020;
Dudok de Wit et al.~2020). Their origin, however, remains elusive.

The structure of the paper entails observations and
data analysis in Section 2, discussions in Section 3,
and conclusions in Section 4. 

\section{	Observations and Data Analysis 	}

\subsection{	Previous Studies of Solar Flare Hard X-rays 	}

In Table 1 we compile power law slopes $\alpha_{\tau}$ of 
published waiting time distributions of solar flares, which we 
briefly describe in turn.

The study of Pearce et al.~(1993) determines a total of 8319 
waiting times (within a time range of $\tau=1-60$ min during
the observation period of 1980-1985), using data from the 
{\sl Hard X-ray Burst Spectrometer (HXRBS)} 
onboard the {\sl Solar Maximum Mission (SMM)}. 
Similarly, Biesecker (1994) detects a total of 6596 waiting times,
using {\sl Burst And Transient Source Experiment (BATSE)} onboard 
the {\sl Compton Gamma Ray Observatory (CGRO)}. 
A much smaller data set with 182 waiting times was obtained from
the {\sl Wide Angle Telescope for Cosmic Hard X-rays (WATCH)} onboard 
the GRANAT satellite (Crosby 1996).  
More extended statistics in hard X-ray wavelengths was gathered from 
the {\sl International Cometary Explorer (ICE)} onboard the
{\sl International Sun/Earth Explorer (ISEE-3)} 
(Wheatland et al.~1998), HXRBS/SMM (Aschwanden and McTiernan 2010),
BATSE/CGRO (Grigolini et al.~2002; Aschwanden and McTiernan 2010),
and the {\sl Ramaty High Energy Solar Spectroscopic Imager (RHESSI)}
(Aschwanden and McTiernan 2010). 

These studies are all based on hard X-ray solar flare catalogs, where
waiting times are found within a range of $\tau\approx 0.01-1000$ hrs
(see time ranges in Table 1), detected above a typical flux threshold of 
$F_{\rm HXR} \gapprox 100$ cts s$^{-1}$ at hard X-ray energies of 
$E_{\rm HXR} \gapprox 25$ keV. Most waiting time distributions 
are self-similar (or scale-free) and consequently exhibit a power 
law scaling. We find only two exceptions, one is an exponential case, 
and another is a double-exponential case, see column 5 in Table 1.
Thus, most of the waiting time distributions of hard X-ray data
can be fitted with a power law function, with a typical slope of
$\alpha_{\tau} \approx 2.0$ (see column 6 in Table 1),
except for two cases with relatively small waiting time ranges.
These two cases exhibit unusual low values of $\alpha_{\tau}
\lapprox 0.8$ (Pearce et al.~1993; Crobsy 1996), because the 
fitted range is too narrow and suffers from incomplete sampling 
(Aschwanden and McTiernan 2010). Hence we discard these two 
dubious cases in the following analysis. 

\subsection{	Previous Studies of Solar Flare Soft X-rays 	}

More waiting time distributions of solar flares were sampled in
soft X-ray wavelengths, making use of the 1-8 \ang\ flux data
observed with the {\sl Geostationary Orbiting Earth Satellite 
(GOES)}, which yields an uninterrupted time series of up to 47
years (from 1974 to present). 
Subsets of GOES time series in different year ranges
were analyzed by Boffetta et al.~(1999),
Wheatland (2000a, 2003), and Lepreti et al.~(2001).
These data are ideal for
waiting time statistics, since the duty cycle of GOES data
is very high (94\%) thanks to the geostationary orbit.
The number of solar flares above a threshold of the C-class
level ($10^{-6}$ W m$^{-2}$) amounts to 35,221 events,
while a deep survey with automated flare detection and 
with $\approx 10$ times higher sensitivity revealed a total
of 338,661 events (Aschwanden and Freeland 2012).
This data set provides the largest statistics of waiting 
times and is investigated in Section 2.5. 
All waiting time distributions could be fitted with a
{\sl ``thresholded''} power law distribution (Aschwanden 2015),
also called Pareto function (Eq.~6), or Lomax function
(Lomax 1954; Hosking and Wallis 1987), which yields here
power law slopes in the range of $\alpha_{\tau}
\approx 1.4-3.2$ (see column 6 in Table 1). 
Note that these results encompass a large range of power 
law slopes that needs to be explained by any theoretical model.
Moreover, the formal errors or uncertatinties in the power law
slopes is typically $\sigma_{\tau} \approx 0.1-0.2$,
which is much smaller than the spread of the slope values
$\alpha_{\tau}\approx 1.4-3.2$, and thus cannot be
explained with a single theoretical constant.

\subsection{	Previous Studies of Coronal Mass Ejections 	}

There is a general consensus now that solar flare events
and {\sl coronal mass ejections (CMEs)} have very high 
mutual association rates, especially for eruptive flare events,
so that the event statistics of one event group can be used 
as a proxy of the other event group. Hence we expect that the
waiting time distributions of solar flares and CMEs are
similar. Wheatland (2003) sampled the waiting time distributions
of CMEs based on {\sl Large Angle Solar Spectrometric
Coronagraph (LASCO)} data from the {\sl Solar and Heliospheric
Observatory (SOHO)} spacecraft. These waiting time distributions
could be fitted with thresholded power law (or Pareto)
functions also, with power law slopes in the range of
$\alpha_{\tau} \approx 1.9-3.0$, sampled in different
time phases of the solar cycle (see column 6 in Table 1).
It turns out that the flare waiting times produce similar
power law slopes $\alpha_\tau$ as the CME waiting times, 
if one compares data from the same phase of the solar cycle. 

\subsection{	Solar Cycle Dependence of Waiting Times 	}

Wheatland (2003) examined the distribution of waiting times 
between subsequent CMEs in the LASCO CME catalog for the years 
1996-2001 and found a power-law slope of 
$\alpha_{\tau} \approx 2.36\pm0.11$ for large 
waiting times (at $\ge$10 hours). Wheatland (2003)
noted that the power law index of the waiting-time 
distribution varies with the solar cycle: for the years
1996-1998 (a period of low activity), the power-law slope is 
$\alpha_{\tau}\approx 1.86\pm 0.14$, and for
the years 1999-2001 (a period of higher activity), the slope is 
$\alpha_{\tau}\approx 2.98\pm 0.20$. Wheatland (2003)
concluded that the observed CME waiting time distribution, 
and its variation with the solar cycle, may be understood in 
terms of CMEs occurring as a time-dependent Poisson process.
The same can be said for solar flares, since flares and CMEs
are highly correlated.

In order to prove this hypothesis, we investigate the
functional relationship between the waiting time 
power law slope $\alpha_{\tau}$ and the
level of solar activity, which we quantify by the
sunspot number SN. We list the mean annual sunspot number 
(from the Royal Observatory in Belgium, www.sidc.be/silso/) 
in Table 2 for the years of 1974-2012. Then we average
the annual sunspot numbers over the range of years
that correspond to each data set given in Table 1:
1980-1989 (HXRBS/SMM; Aschwanden and McTiernan 2010);
1991-1993 (BATSE/CGRO; Aschwanden and McTiernan 2010);
1991-2000 (BATSE/CGRO; Grigolini et al.~2002);
2002-2008 (RHESSI; Aschwanden and McTiernan 2010);
1975-1999 (GOES; Boffetta et al.~1999; Wheatland 2000a; 
                 Lepreti et al.~2001),
1996-2001 (GOES, LASCO/SOHO; Wheatland 2003);
1996-1998 (GOES, LASCO/SOHO; Wheatland 2003);
1999-2001 (GOES, LASCO/SOHO; Wheatland 2003);
2018	  (PSP; Dudok de Wit et al.~2020). 
Then we plot the power law slopes $\alpha_{\tau}$
as a function of the annually averaged sunspot numbers SN
in Fig.~1 and find a linear relationship of
\begin{equation}
	\alpha_{\tau} = 1.38\ +\ 0.01\ \times\ {\rm SN} \ , 
\end{equation}
as confirmed by the high value of the Pearson's 
cross-correlation coefficient (CCC=0.987). 
The sunspot number varies from SN$\approx$40 during
the solar cycle minimum to SN$\approx$160 during the
solar cycle maximum. 
Interestingly, this result is robust in the sense that it 
predicts the same linear relationship even when different
event detection thresholds, different phenomena 
(flares, CMEs, solar wind switchbacks), and different wavelengths
(hard X-rays, soft X-rays) are used. This empirical
linear relationship explains a variety of observed
power law slopes for waiting times, covering a range of 
$\alpha_{\tau} \approx 1.4-3.2$ (Fig.~1).

\subsection{	Annual GOES Waiting Time Statistics	}

The published power law slopes mentioned in the
previous Section mostly cover multi-year ranges.
In the following task we examine the degree of correlation 
between the annual power law slopes of waiting times and 
the annual sunspot numbers from the entire 47-year data set 
of GOES flares. We break the data sets down into 39 annual 
groups (from 1974 to 2012),
for the same time epoch as we performed automated flare
detection in a previous study (Aschwanden and Freeland 2012).
For each of the 39 data sets we fit a thresholded
power law (or Pareto-type) distribution) (Eq.~6), parameterized
with $(N_0, \tau_0, \alpha_{\tau m})$. 

The published power law slopes $\alpha_\tau$ are usually fitted 
in the inertial range $(x_0, x_2)$ and have a mean value that
corresponds to the (logarithmic) midpoint $x_3 \approx (x_0+x_2)/2$ 
of the waiting time distribution. In contrast, the fitting of a
power law function plus an exponential cutoff has a bias to
yield steeper power law slopes $\alpha_{\tau m}$ at the upper end $x_2$,
where the slope is steepest. In order to make the two methods
compatible, we can define an empirical correction factor $q_m$, for
which we find an average value of $q_m \approx 0.62$,
\begin{equation}
	\alpha_\tau = q_m \alpha_{\tau m} 
	            \approx 0.62\ \alpha_{\tau m} \ .
\end{equation}
Applying this correction to the upper limits $\alpha_{\tau m}$
(crosses in Fig.~3a), the corrected power law slopes $\alpha_{\tau}$
(crosses  in Fig.~3b) become fully compatible with the 
published power law values $\alpha_\tau$ (diamonds in 
Figs.~1, 3a, and 3b) based on standard power law fitting
methods in the observed inertial ranges. 

We present the fits of the Pareto distribution functions
(Eq.~6) in Fig.~2 and list the years, the number of events $n_{ev}$,
the power law slope $\alpha_{\tau}$ of the waiting times,
the goodness-of-fit $\chi^2$, the lower $x_0$ and upper
bounds $x_2$ of the fitted range, the decades of the
inertial range $log(x_2/x_0)$, and the annual sunspot
number SN in Table 2. 

Plotting the time evolution of the power law slope 
$\alpha_{\tau}(t)$ in annual steps (Fig.~4a),
we see a solar cycle variation that is similar to
the time variation of the annual sunspot number SN(t)
(Fig.~4b), or the flaring rate per year (Fig.~4c).
Fitting a linear regression between the
two parameters, we find a linear relationship of
$\alpha_{\tau} = 1.442\ +\ 0.009\ \times\ {\rm SN}$
(Fig.~5a), with a correlation coefficient of CCC=0.961. 
This correlation corroborates the result of a
linear relationship between the power law slope and
the sunspot number from previous work (Fig.~1).

The value of the power law slope $\alpha_\tau$,
after the inertial range correction, appears 
not to depend on event detection
thresholds, efficiency of event detection,
and pulse pile-up effects. 
For long-duration flare events, short flares could 
pile-up upon on long tails, violating the
separation of time scales, and possibly steepening
the power law slope of waiting times.
Our automated flare detection is about 10
times more sensitive than the standard NOAA flare
catalog, while the GOES C-class level is often
used as a threshold in statistical studies. 

\subsection{ Switchback Events detected with the Parker Solar Probe }

More recently, the Parker Solar Probe (PSP) 
(Fox et al. 2015) has revealed the omnipresence 
of so-called switchback events in Alfv\'enic 
solar wind streams within a distance of 50 
solar radii from the Sun. These events show up 
as sudden deflections of the (otherwise mostly 
radial) magnetic field of the young solar wind. 
Because switchbacks can easily be identified as 
discrete events, they offer yet another 
opportunity for investigating how waiting time 
distributions are influenced by solar activity.

Dudok de Wit et al.~(2020) analyzed a time 
interval that is centered on the first perihelion 
pass of November 6, 2018 and runs from November 
1st to November 10, 2018. The vector magnetic field
onboard PSP is measured by the MAG magnetometer 
from the FIELDS instrument suite (Bale et al.~2016).
The waiting time and residence time scales of 
switchbacks occur predominantly in the inertial 
range of the solar wind: the sampled waiting time 
distribution covers a range of $\tau \approx
10^{-6}-10^2$ hours. The power law slope varies 
in a range of $\alpha_\tau \approx 1.4 - 1.6$
(Dudok de Wit et al.~2020).

Here we revisit these results by considering a new data
set of $n_{ev}$=19,543 individual switchback events 
that have been automatically selected by feature 
recognition using magnetic field data sampled at
0.1 s.  Switchback events are defined here as a step 
function in the deflection of the magnetic field 
with respect to the orientation of the Parker 
spiral. These events are required to be aligned 
for some duration before and after the event. 
Waiting times with $\tau < 10$ s are believed to be 
biased by smaller events, and waiting times with 
$\tau < 2$ s are increasingly polluted by ion 
cyclotron waves and are discarded anyway. Let us 
stress that unlike with flare events, in which 
large flares can be followed by smaller ones, no 
switchbacks can occur during long switchback events. 
For that reason the long tails of the distributions 
may differ, because the separation of time scales
(waiting times, flare durations) is violated.

We show the waiting time distribution in Fig.~(6), 
fitted with two theoretical models (see Section 3.1),
with (i) the Pareto distribution that is close to a 
power law function at large waiting times (Fig.~6a), 
and (ii) with a Pareto plus an exponential cutoff 
(Fig.~6b).  The latter model yields a superior fit 
(Fig.~6b), with a power law slope of 
$\alpha_{\tau} = 1.21 \pm 0.01$ 
and a goodness-of-fit $\chi=2.21$, which indicates 
that a power law with an exponential cutoff is a 
more realistic description of the waiting time 
distribution.

\section{	Discussion 		}

\subsection{	Pareto-Type Waiting Time Distribution Function }

The statistics of waiting times bears information that enables us 
to discriminate between two statistical distribution functions: 
(i) random processes with Poissonian noise, and 
(ii) clustering of events within individual intervals 
(such as Omori's law for earthquakes, which exhibits 
precursors and aftershock events during a major complex 
earthquake event). A Poissonian process in the time domain 
is a sequence of randomly distributed and thus statistically 
independent events, producing a waiting time distribution 
of time intervals $\tau$ that follows an exponential function,
\begin{equation}
	N(\tau) \ d\tau = \lambda_0\  
	\exp{\left( - {\lambda_0 \ \tau} \right) } \ d\tau \ ,
\end{equation}
where $\lambda_0 = 1/\tau_0$ is the mean flaring rate
or mean reciprocal waiting time,
and $N(\tau)$ is the probability density function
(or differential occurrence frequency). 
Therefore, if individual events 
are produced by a physical random process, the occurrence
frequency of waiting times should follow such an exponential-like
function (Eq.~3).

In reality, however, almost all observed waiting time
distributions of solar flare events exhibit a power law-like 
distribution for the probability function $N(\tau)$, 
rather than an exponential function. In an attempt to 
match this observational constraint, a non-stationary
flare occurrence rate $\lambda(t)$ was defined that 
varies as a function of time in piece-wise time intervals or
Bayesian blocks (Wheatland et al.~1998;
Wheatland 2000a, 2001, 2003).
Some examples of such non-stationary Poisson processes
are given in Aschwanden and McTiernan (2010) and
Aschwanden (2011a, Section 5.2). For instance, it 
includes an exponentially growing (or decaying) flare
rate $\lambda(t)$ that produces a Pareto-type distribution 
for the waiting time distribution with a power law slope of 3,
\begin{equation}
	N(\tau ) \ d\tau = {2 \lambda_0 \over
	(1 + \lambda_0 \tau)^3} \ d\tau \ .
\end{equation}
It includes also a flare rate that varies highly intermittently
in form of $\delta$-functions, which produces a Pareto
distribution also, but with a different power law 
slope of 2, 
\begin{equation}
	N(\tau) \ d\tau = {\lambda_0 \over
	(1 + \lambda_0 \tau)^2} \ d\tau \ ,
\end{equation}
In order to generalize these two solutions (Eqs.~4 and 5)
into a single function, we can define a Pareto-type 
distribution with a variable power law slope
$\alpha_{\tau}$, as parameterized in Eqs.~(1) and (2),
\begin{equation}
	N(\tau) \ d\tau = N_0 (\tau_0 + \tau)^{-\alpha_{\tau}} 
	\ d\tau \ .
\end{equation}

A more general expression of the waiting time distribution 
that includes finite system size effects, in terms of an
exponential-like cutoff near the longest waiting time
intervals $\tau \lapprox \tau_e$, 
\begin{equation}
        N(\tau) d\tau 
	= N_0 \left( \tau_0 + \tau \right)^{-\alpha_{\tau m}}
        \exp{ \left( - {\tau \over \tau_e}\right) } \ d\tau \ .
\end{equation} 
which yields a superior fit, as shown in the case of solar
wind switchbacks (Fig.~6b). 
This equation fully describes the observed waiting
time distributions (as shown in Figs.~2 and 6), expressed by five 
parameters ($N_0, \tau_0, \tau_e, \alpha_{\tau m}$, SN). 
The mean waiting time $\tau_0$, approximately
represents the lower bound of the inertial range and
separates the range of incompletely sampled events 
(see Fig.~6b) from 
the scale-free power law range of completely sampled events.
The waiting time $\tau_e$ demarcates the e-folding cutoff
due to finite system size effects. $\tau_0$ is an offset 
that depends on the 
instrumental  sensitivity, the flux threshold in the automated 
detection of waiting times, and pulse-pileup effects. 
For long-duration flare events, short flares may
break up long waiting times into smaller waiting times,
which steepens the power law slope.
Since the sunspot number SN is an observable that is known 
from earlier centuries up to today (Table 2), there are only four 
free variables left ($N_0, \tau_0, \tau_e, \alpha_{\tau m}$), which can 
be derived empirically by fitting Eqs.~(7) to an
observed data set. Note that the maximum power law slope 
$\alpha_{\tau m}$ relates to the mean power law slope
$\alpha_{\tau}$ by a correction factor given in Eq.~(2).

The fact that most waiting time distributions exhibit a
power law-like function, rather than an exponential-like
function, clearly requires a non-stationary Poisson process
(Wheatland et al;.~1998; Wheatland 2000a, 2001), which 
implies that the flaring rate $\lambda(t)$ has a substantial time
variability. The time variability of the mean flaring rate
has been found to be highly correlated with the annual
sunspot number (Fig.~4), and thus is modulated
by Hale's magnetic cycle of solar activity. 
Apparently, the mean waiting time between solar flares depends 
on the global magnetic flux (quantified by the sunspot number), 
but not on the flare size 
(or peak count rate) according to observations, in contrast to
theoretical expectations of energy storage models 
(Rosner and Vaiana 1978; Wheatland 2000b).

\subsection{	Flare Model of Waiting Times 		}

All fitted waiting time distributions shown in Fig.~2 are
modulated by annual variations of the solar cycle.
On shorter (than annual) time scales, the flaring rate varies
also, as statistics based on Bayesian-block decomposition
reveal (Wheatland et al.~1998; Wheatland 2000b, 2001, 2003; 
Wheatland and Litvinenko 2002). A flare model that possibly
could explain the waiting time distribution function was
proposed (Wheatland and Litvinenko 2002; 
Wheatland and Craig 2006), based on the
assumptions of: (i) Alfv\'enic time scale for crossing the
magnetic reconnection region, (ii) 2-D geometry of reconnetion
region (or separatrix); and (iii) correlation of flare energy 
build-up (or storage) and flare waiting time. 
This model, however, is not consistent
with observations, which show no correlation between flare
sizes and flare waiting times (Crosby 1996; Wheatland 2000b;
Georgoulis et al.~2001), not even between subsequent
flares of the same active region (Crosby 1996; Wheatland 2000b).
Consequently, such theoretical flare energy storage models 
(Rosner and Vaiana 1978; Lu 1995) have been abandoned.
A more likely model involves interchange reconnection
between coronal loops and open magnetic fields (Zank et al.~2020).

\subsection{	Self-Organized Criticality Models	}

The {\sl fractal-diffusive avalanche model of 
a slowly-driven self-organized criticality (FD-SOC) system}
(Aschwanden 2012, 2014; Aschwanden and Freeland 2012;
Aschwanden et al.~2016), expanded from the original version
of Bak et al.~(1987, 1988), is based
on a scale-free (power law) size distribution function
of avalanche (or flare) length scales $L$,
\begin{equation}
	N(L)\ dL \propto L^{-d}\ dL \ ,
\end{equation}
with $d$ the Euclidean spatial dimension (which can have
values of d=1, 2, or 3). This reciprocal relationship
between the spatial size $L$ of a switchback structure
and the occurrence frequency $N(L)$ is visualized
in Fig.~8, for the case of a space-filling avalanche mechanism,
but it holds for rare events in terms of relative 
probabilities also. 

The transport process of an avalanche is described by
classical diffusion according to the FD-SOC model, 
which obeys the scaling law,
\begin{equation}
	L \propto T^{\beta/2} \ ,
\end{equation}
with $\beta=1$ for classical diffusion. Substituting
the length scale $L \propto T^{\beta/2}$ with  
the duration $T$ of an avalanche event, using Eq.~(8-9)
and the derivative $dL/dT = T^{\beta/2-1}$, predicts  
a power law distribution function for the size distribution
of time durations $T$,
\begin{equation}
	N(T)\ dT = N(T[L]) \left({dL \over dT}\right) \ dT 
		 = T^{-[ 1+ (d-1) \beta/2] } dT 
		 = T^{-\alpha_\tau}\ dT
		 \approx T^{-2} dT \ ,
\end{equation}
for $d=3$ and $\beta=1$, defining the waiting time power
law slope $\alpha_\tau$,
\begin{equation}
	\alpha_\tau = 1+ (d-1) \beta/2 = 2  \ .
\end{equation}
We can now estimate the size distribution of 
waiting times by assuming that the avalanche durations
represent upper limits to the waiting times $\tau$ during
flaring time intervals, while the waiting times become
much larger during quiescent time periods. Such a bimodal
size distribution with a power law slope of $\alpha_\tau \lapprox 2$
at short waiting times ($\tau \le \tau_e$), 
and an exponential-like cutoff function
at long waiting times $(\tau \ge \tau_e)$, is depicted in Fig.~7, 
\begin{equation}
	N(\tau )\ d\tau = \left\{ \begin{array}{ll}
	\tau^{-2}                    & {\rm for}\ \tau \ll \tau_e \\ 
	\tau^{-2} \exp(-\tau/\tau_e) & {\rm for}\ \tau \gapprox \tau_e \\ 
		  \end{array} \right. \ .
\end{equation} 
Thus, this FD-SOC model predicts a power law with a slope of
$\alpha_\tau \approx 2.0$ in the inertial range, and a steepening 
cutoff function at longer waiting times. Observationally, we find
a range of $\alpha_\tau \approx 1.4-3.2$, varying as a function
of the solar activity, but the solar cycle modulation is not
quantified in any SOC model (Aschwanden 2019b). 

\subsection{	MHD Turbulence Processes		}

Boffetta et al.~(1999) argue that the statistics of solar
flare (laminar or quiescent) waiting times indicate a
physical process with complex dynamics with long correlation 
times, such as in chaotic models, in contradiction to 
stationary SOC models that predict Poisson-like 
statistics. They consider chaotic models that include the 
destabilization of the laminar phases and subsequent 
restabilization due to nonlinear dynamics, as invoked in
their shell model of MHD turbulence.

Similarly, Lepreti et al.~(2001) attribute the origin of the 
observed waiting time distribution to the fact that the physical
process underlying solar flares is statistically self-similar 
in time and is characterized by a certain amount of
``memory''. They find that the power law distribution can be
modeled by a L\'evy function which can explain a power law 
exponent of $\alpha_\tau=3$ (Eq.~4) in the waiting time
distribution. 

Grigolini et al.~(2002) develop a technique called diffusion
entropy method to reproduce the observed waiting time 
distribution function,
which evaluates the entropy of the diffusion process
generated by the time series. Note that classical diffusion
(Eq.~9) has been employed in SOC models (Aschwanden 2012), 
which may be related to the diffusion entropy method, since 
both models produce a similar scaling of short waiting times, 
with power law slopes in the range of $\alpha_\tau \approx 1.4-3.2$ 
(Table 1). The change of the power law index
from $\alpha_\tau > 3$ to $\alpha_\tau < 3$ has been interpreted
in terms of a phase transition from the Gaussian to the
L\'evy basin of attraction (Grigolini et al.~2002).

\subsection{	Switchback events observed with Parker Solar Probe }

The novel phenomenon of the so-called switchback events were
sampled in situ with the PSP in the solar wind at a distance of 
$\approx 36 R_{\odot}=0.166$ AU. Switchback events have
durations of less than 1 s to more than an hour. Hallmarks of
switchback events are reversals in the radial field component
$B_r$ (with respect to the Parker spiral geometry), which can produce 
deflection angles from a few degrees to nearly $180^\circ$
in the fully anti-sunward direction (Dudok de Wit et al.~2020). 
A switchback event can
be quantified either by the magnetic potential energy,
\begin{equation}
	E_p = - {\bf B}\ \cdot\ <{\bf B}> \ , 
\end{equation} 
or by the normalized deflection angle $\mu$,,
\begin{equation}
	z = {1\over 2} ( 1 - \cos{\mu} ) \quad {\rm for}\
	0 \le z \le 1 \ .
\end{equation}
Moreover, we can sample the waiting times $\tau_i = (t_{i+1}-t_i)$
of subsequent events, which resulted into a power law-like
inertial range of $\tau \approx 10-500$ s, an exponential cutoff
at $\tau \approx 500-2000$ s, and an undersampled range at
$\tau \approx 0.1-10$ s (Fig.~6b). The power law slope of the
waiting time distribution, $\alpha_\tau=1.21\pm0.01$ (Fig.~6),
is measured in the year 2018, close to the minimum of the
solar cycle, and follows the same trend as solar flares and
CMEs (Fig.~1). 

This close similarity of the slopes obtained with 
solar flares and with switchbacks is intriguing 
and could be the signature of common drivers. 
However, as of today the origin of switchbacks 
is unclear. They may be generated either locally 
in the upper solar corona or by instabilities such 
as plasma jets occurring much deeper in the corona. 
In addition, there are also several differences 
in the way flares and switchbacks are registered: 
the flaring rate, for example, includes independent 
and sympathetic flares occurring everywhere on the 
solar disc and at the limb, while switchback events 
are recorded at one single point in space only, 
along the satellite orbit. In addition, the impact 
of local solar wind conditions, and the relative 
speed of PSP on the rate of switchback events 
still has to be properly investigated. Therefore, 
while the similarity of the waiting time 
distributions is likely to be deeply rooted in 
the underlying physical processes, it is premature 
to conclude about the connection between the two 
types of events.

Long-term memories, expressed by the residence time of switchback 
events are then expected to scale with the flare or CME duration $T$,
which is predicted from SOC models to follow a power law distribution
function of $N(T) \propto T^{-2}$ (Eq.~10), and a proportional
distribution of $N(\tau ) \propto \tau^{-2}$ for short waiting
times (Eq.~12). 
{\sl Both the waiting time $\tau$ and the residence time $T$ distribution 
of these switchback deflections tend to follow a power law and 
are remarkably similar} (Dudok de Wit et al.~2020).
{\sl The long memory we observe is most likely associated with the 
strong spatial connection between adjacent magnetic flux tubes and 
their common photospheric footpoints} (Dudok de Wit et al.~2020). 
Consequently, it has been proposed that switchback events are modulated
by impulsive flare (or CME) events in the lower corona 
(Roberts et al.~2018; Tenerani et al.~2020; Zank
et al.~2020).

\section{ 	Conclusions	}

In this study we investigate the statistics of waiting time
distributions of solar flares, CMEs, and solar wind switchback
events. The motivation for this type of analysis method is
the diagnostics of stationary and non-stationary Poissonian 
random processes, SOC systems, and
MHD turbulence systems. The observational analysis is very
simple, since only an event catalog with the starting
times $t_i$ of the events is necessary to sample waiting
times $\tau=(t_{i+1}-t_i)$. We obtain the following results:

\begin{enumerate}

\item{Using the statistics of hard X-ray solar flares
(using flare catalogs from HXRBS/SMM, BATSE/CGRO, WATCH, ICE/ISEE-3,
RHESSI) we find power law distribution functions with slopes
in the range of $\alpha_\tau \approx 1.5-3.2$, following
a linear regression fit of $\alpha_{\tau m}=1.38 \pm 0.01
\ \times\ $ SN and a cross-correlation coefficient of CCC=0.987
(Fig.~1).
This trend clearly indicates that the waiting time power law slope
$\alpha_\tau$ 
is foremost correlated with the sunspot number (or the flaring rate),
which is fully consistent with previous findings (Wheatland
and Litvinenko 2002; Wheatland 2003).}

\item{Using the statistics of soft X-ray flares, sampled by GOES
over 47 years
in annual intervals, but with a 10 times higher sensitivity,
we perform fits with a Pareto-type distribution function (Fig.~2), 
which consists of an inertial (power law) range, an exponential
cutoff range, and a range of under-sampling. The fits clearly
show power law slopes that are modulated by the four solar cycles,
strongly correlated with the annual sunspot number and the 
annual flaring rate (Figs.~4, 5), consistent with the hard X-ray
flare results.}

\item{We sample 19,452 magnetic field switchback events from
data observed with the Parker Solar Probe and find a power law
slope of $\alpha_\tau=1.21\pm 0.01$. A theoretical value of
$\alpha_\tau = 2.0$ is predicted for short waiting
times ($\tau \lapprox 500$ s) 
by a self-organized criticality model during contigous flaring time
episodes, while an exponentially droping cutoff is expected
for long waiting times ($\tau \approx 500-2000$ s). Hence
we propose that a realistic waiting time distribution
contains a power law part for short waiting times and an
exponential part for long waiting times (Eq.~12), which reflects
the duality between flaring and quiescent episodes,
corresponding to a combination of non-stationary and stationary 
Poissonian components, which can be modeled with piece-wise Bayesian
time intervals (e.g., Wheatland and Litvinenko 2002).}

\item{Although the four theoretical models discussed here
(non-stationary Poissonian, stationary Poissonian, SOC, and
MHD turbulence) can all explain some partial aspects of the
obseved waiting time distribution functions (exponential,
power law), none of them predicts the most dominating parameter,
namely the time-dependent modulation of the magnetic solar cycle, 
which can be modeled in terms of the flaring rate or sunspot number. 
This result suggests that the variability observed in the solar wind
is modulated by flares and CMEs originating in the lower corona,
rather than in localized heliospheric sources.
Further conceptual uncertainties in data modeling include
under-sampling and thresholding in automated event detection,
pulse pile-up of small events in the wake of larger events,
Bayesian interval selection, distinction of flaring and
quiescent episodes, and exponential cutoff at maximum waiting
times. A Pareto-type function with an exponential cutoff appears
to be the most appropriate choice to fit the observed waiting time
distributions.}

\end {enumerate}
  
Waiting time statistics has been applied to many nonlinear
phenomena. While we deal here with three phenomena only
(solar flares, CMEs, solar wind switchback events), other
analyzed data include earthquakes (Omori's law), auroral
emission, substorms in magnetosphere, solar radio bursts,
stellar flares (Aschwanden 2019a; Aschwanden and G\"udel 2021), 
black hole accretion disks (see Aschwanden 2011a, 2016, 
and references therein). 

\vskip1cm
{\sl Acknowledgements:}
Part of the work was supported by NASA contract NNG04EA00C of the
SDO/AIA instrument and NNG09FA40C of the IRIS instrument. 

\clearpage

\section*{ References }  

\def\ref#1{\par\noindent\hangindent1cm {#1}} 

\ref{Aschwanden, M.J. and McTiernan, J.M. 2010,
	{\sl Reconciliation of waiting time statistics of 
	solar flares observed in hard X-rays},
 	ApJ 717, 683}
\ref{Aschwanden, M.J. 2011a
 	{\sl Self-Organized Criticality in Astrophysics. 
	The Statistics of Nonlinear Processes in the Universe},
 	ISBN 978-3-642-15000-5, Springer-Praxis: New York, 416p.}
\ref{Aschwanden, M.J. 2011b,
 	{\sl The state of self-organized criticality of the Sun 
	during the last 3 solar cycles. I. Observations},
 	SoPh 274, 99}
\ref{Aschwanden, M.J. 2012,
 	{\sl A statistical fractal-diffusive avalanche model of 
	a slowly-driven self-organized criticality system}
 	A\&A 539:A2.}
\ref{Aschwanden, M.J. and Freeland, S.M. 2012,
	{\sl Automated solar flare statistics in soft X-rays over
	37 years of GOES observations: The invariance of SOC
	during 3 solar cycles},
	ApJ 754:112.}
\ref{Aschwanden, M.J. 2014,
	{\sl A macroscopic description of self-organized systems
	and astrophysical applications},
	ApJ 782, 54}
\ref{Aschwanden, M.J. 2015,
	{\sl Thresholded power law size distributions of instabilities
	in astrophysics}, 
	ApJ 814:19.}
\ref{Aschwanden, M.J. 2016,
	{\sl 25 Years of SOC: Solar and astrophysics}
	SSRv 198:47.}
\ref{Aschwanden, M.J. 2019a,
	{\sl Self-organized criticality in solar and stellar flares:
	Are extreme events scale-free ?}
	ApJ 880, 105.}
\ref{Aschwanden, M.J. 2019b,
	{\sl Nonstationary fast-diven, self-organized criticality in
	solar flares},
	ApJ 887:57}
\ref{Aschwanden, M.J. 2020,
	{\sl Global energetics of solar flares. XII. Energy scaling laws},
	ApJ (in press).}
\ref{Aschwanden M.J. and G\"udel, M. 2021,
	{\sl Self-organized criticality in stellar flares},
	(subm.)}
\ref{Bak, P., Tang, C., and Wiesenfeld, K. 1987,
        {\sl Self-organized criticality - An explanation of 1/f noise},
        Physical Review Lett. {\bf 59/27}, 381-384.}
\ref{Bak, P., Tang, C., and Wiesenfeld, K. 1988,
        {\sl Self-organized criticality},
        Physical Rev. A {\bf 38/1}, 364-374.}
\ref{Bale, S.D., Goetz, K., Wygant, J.R. 
	et al.~2016,
	{\sl The FIELDS instrument suite for Solar
	Probe Plus},
	SSRv 204, 49}
\ref{Biesecker, D.A. 1994
 	{\sl On the occurrence of solar flares observed with the 
	burst and transient source experiment}
 	PhD Thesis, University of New Hampshire}
\ref{Boffetta, G., Carbone, V., Giuliani, P., Veltri, P., 
	and Vulpiani, A. 1999,
 	{\sl Power Laws in Solar Flares: Self-Organized Criticality 
	or Turbulence?}
 	Phys.Rev.Lett. 83, 4662}
\ref{Crosby, N.B., Georgoulis, M., and Villmer, N. 1996,
	{\sl A comparison between the WATCH flare data 
	statistical properities and predictions of the
	statistical flare model},
	in Proc. 8th SOHO Workshop (eds. J.C. Vial and 
	B. Kaldeich-Schuermann), ESA 446, Estec Nordwijk, p.247}
\ref{Dudok de Wit, P., Krasnoselskikh, V.V., Bale, S.D., et al. 2020,
	{\sl Switchbacks in the near-Sun magnetic field. Long
	memory and impact on the turbulence cascade},
	ApJSS  246:39}
\ref{Fox, N.J., Velli, M.C., Bale, S.D., Decker, R., Driesman, A., 
	Howard et al. 2016,
 	{\sl The Solar Probe Plus Mission: Humanity's First Visit to 
	Our Star},
 	Space Science Reviews, 204, 7.}
\ref{Georgoulis, M.K., Vilmer, N., Croby, N.B. 2001,2
 	{\sl A Comparison Between Statistical Properties of Solar 
	X-Ray Flares and Avalanche Predictions in Cellular Automata 
	Statistical Flare Models},
 	A\&A 367, 326}
\ref{Grigolini, P., Leddon,D., and Scafetta,N. 2002,
 	{\sl Diffusion entropy and waiting time statistics of hard 
	X-ray solar flares},
 	Phys.Rev.Lett E, 65/4. id. 046203}
\ref{Horbury, S., Woolley, T., Laker R., Matteini, L. et al. 2020
	{\sl Sharp Alfvenic impulses in the Near-Sun solar wind},
	ApJSS 246:45}
\ref{Hosking, J.M.R. and Wallis, J.R. 1987,
	Technometrics 29, 339.}
\ref{Krasnoselskikh, V., Larosa, A., Agapitov, O., Dudok de Wit, T.,
	2020,
	{\sl Localized magnetic Field structures and their 
	boundaries in the near-Sun solar wind from Parker
	Solar Probe measurements},
	ApJ 893, 93.}
\ref{Lepreti, F., Carbone, V., and Veltri,P. 2001,
 	{\sl Solar flare waiting time distribution: varying-rate 
	Poisson or Levy function?}
 	ApJ 555, L133}
\ref{Lomax, K.S. 1954, 
	{\sl }, J. Am. Stat. Assoc. 49, 847}
\ref{Lu, E.T. 1995, 
 	{\sl Constraints on energy storage and release models 
	for astrophysical transients and solar flares}, 
	ApJ 447, 416.}
\ref{Mozer, F.S., Agapitov, O.V., Bale, S.D., Bonnell, J.W. et al. 2020,
	{\sl Switchbacks in the solar magnetic field: Their evolution,
	their contentk, and their effects on the plasma},
	ApJSS 246:68} 
\ref{Pearce, G., Rowe, A.K., and Yeung, J. 1993,
	{\sl A statistical analysis of hard X-ray solar flares}.
	Astrophys. Space Science 208, 99.}
\ref{Roberts M.A., Uritsky, V.M., DeVore, C.R., and Karpen,J.T. 
	2018,
	{\sl Simulated encounters of the Parker Solar Probe 
	with a Coronal-hole Jet},
	ApJ 866, 14}
\ref{Rosner, R., and Vaiana, G.S. 1978,
 	{\sl Cosmic flare transients: constraints upon models for 
	energy storage and release derived from the event frequency 
	distribution}, ApJ 222, 1104}
\ref{Tenerani, A., Velli, M., Matteini, L., et al. 2020,
	{\sl Magnetic Field Kinks and Folds in the Solar Wind},
	ApJS 246, 32}
\ref{Wheatland, M.S., Sturrock,P.A., and McTiernan,J.M. 1998,
 	{\sl The waiting-time distribution of solar flare hard X-rays}
 	ApJ 509, 448-455.}
\ref{Wheatland,M.S. 2000a,
	{\sl The origin of the solar flare waiting-time distribution},
	ApJ 536, L109.}
\ref{Wheatland,M.S. 2000b,
 	{\sl Do solar flares exhibit an internal-size relationship},
	SoPh 191, 381-389.}
\ref{Wheatland, M.S. 2001,
 	{\sl Rates of flaring in individual active regions},
 	SoPh 203, 87-106.}
\ref{Wheatland, M.S. and Litvinenko, Y.E. 2002,
 	{\sl Understanding Solar Flare Waiting-Time Distributions},
 	SoPh 211, 255-274.}
\ref{Wheatland, M.S. 2003,
 	{\sl The Coronal Mass Ejection Waiting-Time Distribution}
 	SoPh 214, 361-373.}
\ref{Wheatland, M.S., and Craig, I.J.D. 2006,
 	{\sl Including Flare Sympathy in a Model for Solar Flare 
	Statistics},
 	SoPh 238, 73-86.}
\ref{Zank, G.P., Nakanotani, M., Zhao, L.L.,
	Adhikari, L., and Kasper, J., 2020,
	{\sl The origin of switchbacks in the
	solar corona},
	ApJ 903:1}

\clearpage

\begin{table}[t]
\begin{center}
\normalsize
\caption{Waiting time distributions measured from solar flares
hard X-ray events, soft X-ray events, coronal mass ejections,
and radio bursts. The waiting time distribution (WTD) functions
are abbreviated as: PL=powerlaw, E=exponential, PE=powerlaw with
exponential cutoff, DE=double exponential.}
\medskip
\begin{tabular}{llrllll}
\hline
Observations    &Observations     &Number         &Waiting        &WTD    &Powerlaw       &References\\
year of events & spacecraft or    &range          &time           &       &               &          \\
             & instrument      &               &$\tau    $     &       &$\alpha_{\tau}$ &     \\
\hline
\hline
1980-1985	& HXRBS/SMM       & 8319          & $0.01-1$ hrs  & PL    & $0.75\pm0.1$  & Pearce \etal (1993)\\
1991-2000       & BATSE/CGRO      & 6596          & $0.03-7$ hrs  & E     &               & Biesecker (1994)\\
1990-1992       & WATCH/GRANAT    & 182           & $0.17-5$  hrs & PE    & $0.78\pm0.13$ & Crosby (1996)\\
1978-1986       & ICE/ISEE-3      & 6916          & $0.01-20$ hrs & DE    &               & Wheatland \etal (1998)\\
1980-1989       & HXRBS/SMM       & 12,772        & $0.01-500$ hrs& PL    & $2.0$         & Aschwanden \& McTiernan (2010)\\
1991-1993       & BATSE/CGRO      & 4113          & $0.01-200$ hrs& PL    & $2.0$         & Aschwanden \& McTiernan (2010) \\
1991-2000       & BATSE/CGRO      & 7212          & $1-5000$ hrs  & PL    & $2.14\pm0.01$ & Grigolini \etal (2002)\\
2002-2008       & RHESSI          & 11,594        & $2-1000$ hrs  & PL    & $2.0$         & Aschwanden \& McTiernan (2010)\\
\hline
1975-1999       & GOES 1-8 A      & 32,563        & $1-1000$ hrs  & PL    & $2.4\pm0.1$   & Boffetta \etal (1999)\\
1975-1999       & GOES 1-8 A      & 32,563        & $1-1000$ hrs  & PL    & $2.16\pm0.05$ & Wheatland (2000a), Lepreti \etal (2001)\\
1996-2001       & GOES 1-8 A      & 4645          & $1-1000$ hrs  & PL    & $2.26\pm0.11$ & Wheatland (2003)\\

1996-1998       & GOES 1-8 A      & ...           & $1-1000$ hrs  & PL    & $1.75\pm0.08$ & Wheatland (2003)\\
1999-2001       & GOES 1-8 A      & ...           & $1-1000$ hrs  & PL    & $3.04\pm0.19$ & Wheatland (2003)\\
1975-2001       & GOES 1-8 A      & ...           & $1-1000$ hrs  & PL    & $2.2\pm0.1$   & Wheatland and Litvinenko (2002) \\
Solar min       & GOES 1-8 A      & ...           & $1-1000$ hrs  & PL    & $1.4\pm0.1$   & Wheatland and Litvinenko (2002) \\
Solar max       & GOES 1-8 A      & ...           & $1-1000$ hrs  & PL    & $3.2\pm0.3$   & Wheatland and Litvinenko (2002) \\
\hline
1996-2001       & SOHO/LASCO      & 4645          & $1-1000$ hrs  & PL    & $2.36\pm0.11$ & Wheatland (2003)\\
1996-1998       & SOHO/LASCO      & ...           & $1-1000$ hrs  & PL    & $1.86\pm0.14$ & Wheatland (2003)\\
1999-2001       & SOHO/LASCO      & ...           & $1-1000$ hrs  & PL    & $2.98\pm0.20$ & Wheatland (2003)\\
\hline
2018            & PSP             & ...           & $10^{-6}-10^2$ hrs & PL & $1.4-1.6$     & Dudok de Wit et al.(2018)\\
2018            & PSP             & ...           & $10^{-6}-10^2$ hrs & PE & $1.21\pm0.01$ & This work \\
\hline 
\end{tabular}
\end{center}
\end{table}

\begin{table}[t]
\begin{center}
\normalsize
\caption{Annual waiting time distribution slopes and sunspot numbers measured from solar flares}
\medskip
\begin{tabular}{rrrrrrrr}
\hline
Year	& Number    & power law   & best-fit  & lower  & upper & decades & Sunspot \\
	& of events & slope       & chi-square& bound  & bound & $\log(x_2/x_0)$ & number \\
	& $N_{ev}$  & $\alpha$    & $\chi^2$  & $x_0$[hrs] & $x_2$[hrs] & & SN \\
\hline
\hline
1974 &    158 & 1.53$\pm$0.12 &  0.95 &    0.078 &   19.869 & 2.4 &  49.2\\
1975 &   2864 & 1.32$\pm$0.02 &  1.24 &    0.025 &   19.514 & 2.9 &  22.5\\
1976 &   1829 & 1.32$\pm$0.03 &  1.17 &    0.029 &  113.170 & 3.6 &  18.4\\
1977 &   5541 & 1.72$\pm$0.02 &  1.30 &    0.037 &    8.309 & 2.3 &  39.3\\
1978 &  16896 & 2.65$\pm$0.02 &  1.69 &    0.024 &    1.319 & 1.7 & 131.0\\
1979 &  18795 & 3.17$\pm$0.02 &  1.51 &    0.026 &    2.180 & 1.9 & 220.1\\
1980 &  15341 & 3.14$\pm$0.03 &  2.10 &    0.026 &    9.511 & 2.6 & 218.9\\
1981 &  15528 & 3.16$\pm$0.03 &  2.72 &    0.024 &    6.005 & 2.4 & 198.9\\
1982 &  15963 & 2.82$\pm$0.02 &  1.98 &    0.023 &    4.945 & 2.3 & 162.4\\
1983 &   9566 & 2.49$\pm$0.03 &  3.27 &    0.038 &    2.819 & 1.9 &  91.0\\
1984 &   6891 & 1.93$\pm$0.02 &  2.53 &    0.035 &    8.690 & 2.4 &  60.5\\
1985 &   2776 & 1.65$\pm$0.03 &  1.53 &    0.051 &   13.588 & 2.4 &  20.6\\
1986 &   2337 & 1.53$\pm$0.03 &  2.04 &    0.046 &   14.600 & 2.5 &  14.8\\
1987 &   4867 & 1.71$\pm$0.02 &  1.04 &    0.030 &   49.343 & 3.2 &  33.9\\
1988 &  13028 & 2.67$\pm$0.02 &  3.01 &    0.028 &    1.034 & 1.6 & 123.0\\
1989 &  17533 & 3.63$\pm$0.03 &  1.77 &    0.033 &    1.078 & 1.5 & 211.1\\
1990 &  15281 & 3.37$\pm$0.03 &  2.48 &    0.032 &    1.053 & 1.5 & 191.8\\
1991 &  16857 & 3.13$\pm$0.02 &  2.04 &    0.024 &    0.847 & 1.5 & 203.3\\
1992 &  12487 & 2.77$\pm$0.02 &  2.42 &    0.033 &    2.248 & 1.8 & 133.0\\
1993 &   9386 & 2.21$\pm$0.02 &  1.43 &    0.038 &    8.785 & 2.4 &  76.1\\
1994 &   4844 & 1.67$\pm$0.02 &  1.25 &    0.039 &    3.972 & 2.0 &  44.9\\
1995 &   2651 & 1.59$\pm$0.03 &  1.82 &    0.079 &   26.021 & 2.5 &  25.1\\
1996 &   1287 & 1.44$\pm$0.04 &  1.27 &    0.066 &   23.094 & 2.5 &  11.6\\
1997 &   3775 & 1.88$\pm$0.03 &  1.71 &    0.051 &   11.941 & 2.4 &  28.9\\
1998 &  11170 & 2.61$\pm$0.02 &  1.17 &    0.040 &    2.100 & 1.7 &  88.3\\
1999 &  14683 & 2.67$\pm$0.02 &  1.02 &    0.025 &    0.827 & 1.5 & 136.3\\
2000 &  16160 & 3.04$\pm$0.02 &  2.24 &    0.025 &    4.052 & 2.2 & 173.9\\
2001 &  16023 & 2.92$\pm$0.02 &  1.39 &    0.027 &    2.201 & 1.9 & 170.4\\
2002 &  16174 & 3.29$\pm$0.03 &  0.81 &    0.031 &    7.031 & 2.4 & 163.6\\
2003 &  12283 & 2.60$\pm$0.02 &  0.99 &    0.033 &    1.129 & 1.5 &  99.3\\
2004 &   8956 & 2.15$\pm$0.02 &  1.15 &    0.038 &    3.057 & 1.9 &  65.3\\
2005 &   6705 & 1.89$\pm$0.02 &  1.16 &    0.032 &   10.065 & 2.5 &  45.8\\
2006 &   3100 & 1.60$\pm$0.03 &  1.24 &    0.045 &   11.067 & 2.4 &  24.7\\
2007 &   1433 & 1.54$\pm$0.04 &  1.20 &    0.039 &   42.179 & 3.0 &  12.6\\
2008 &    190 & 1.51$\pm$0.11 &  0.96 &    0.110 &  170.120 & 3.2 &   4.2\\
2009 &    531 & 1.60$\pm$0.07 &  1.08 &    0.063 &   74.695 & 3.1 &   4.8\\
2010 &   3660 & 1.79$\pm$0.03 &  1.05 &    0.070 &    7.715 & 2.0 &  24.9\\
2011 &  11270 & 2.56$\pm$0.02 &  1.24 &    0.042 &    2.272 & 1.7 &  80.8\\
2012 &  13106 & 2.44$\pm$0.02 &  1.35 &    0.028 &    1.875 & 1.8 &  84.5\\
\hline
\end{tabular}
\end{center}
\end{table}


\begin{figure}
\centerline{\includegraphics[width=1.0\textwidth]{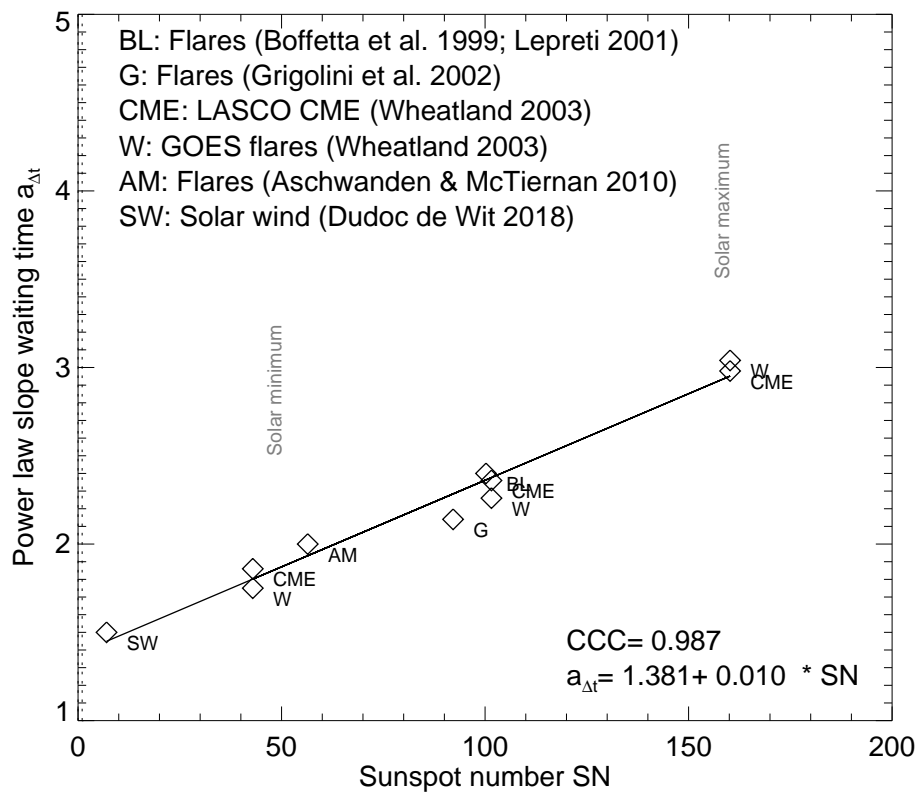}}
\caption{Scatterplot of power law slope $\alpha_{\tau}$
of waiting time distributions versus the sunspot number SN,
with linear regression fit (solid line) and cross-correlation
coefficient CCC.}
\end{figure}

\begin{figure}
\centerline{\includegraphics[width=1.0\textwidth]{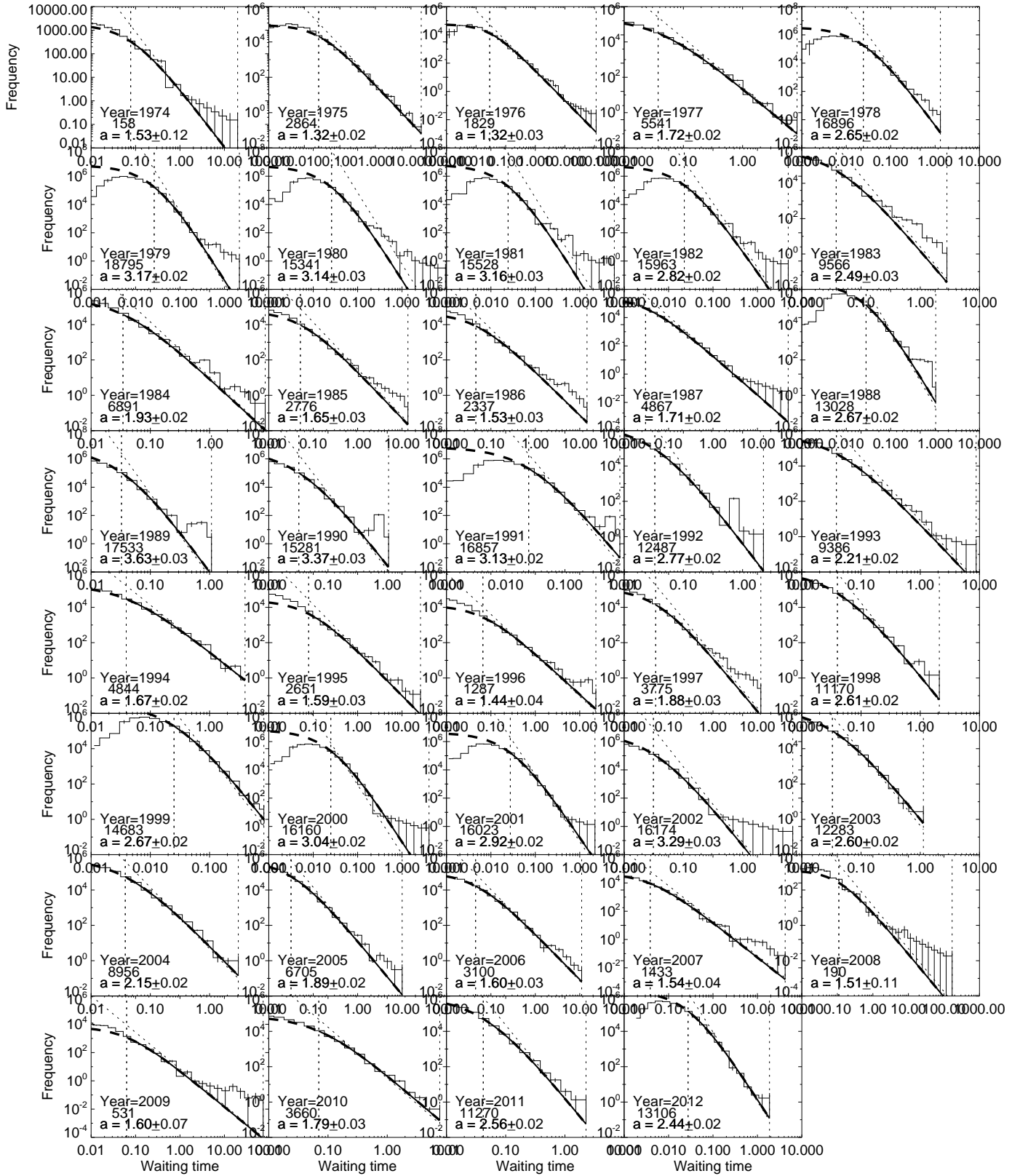}}
\caption{Annual waiting time distributions of GOES 1-8 \ang\
soft X-ray fluxes for the years 1974 to 2012 (histograms),
with least-square fits of Pareto distributions (thick solid 
curve) and corresponding power law slopes $a$  (dotted lines)}
\end{figure}

\begin{figure}
\centerline{\includegraphics[width=1.0\textwidth]{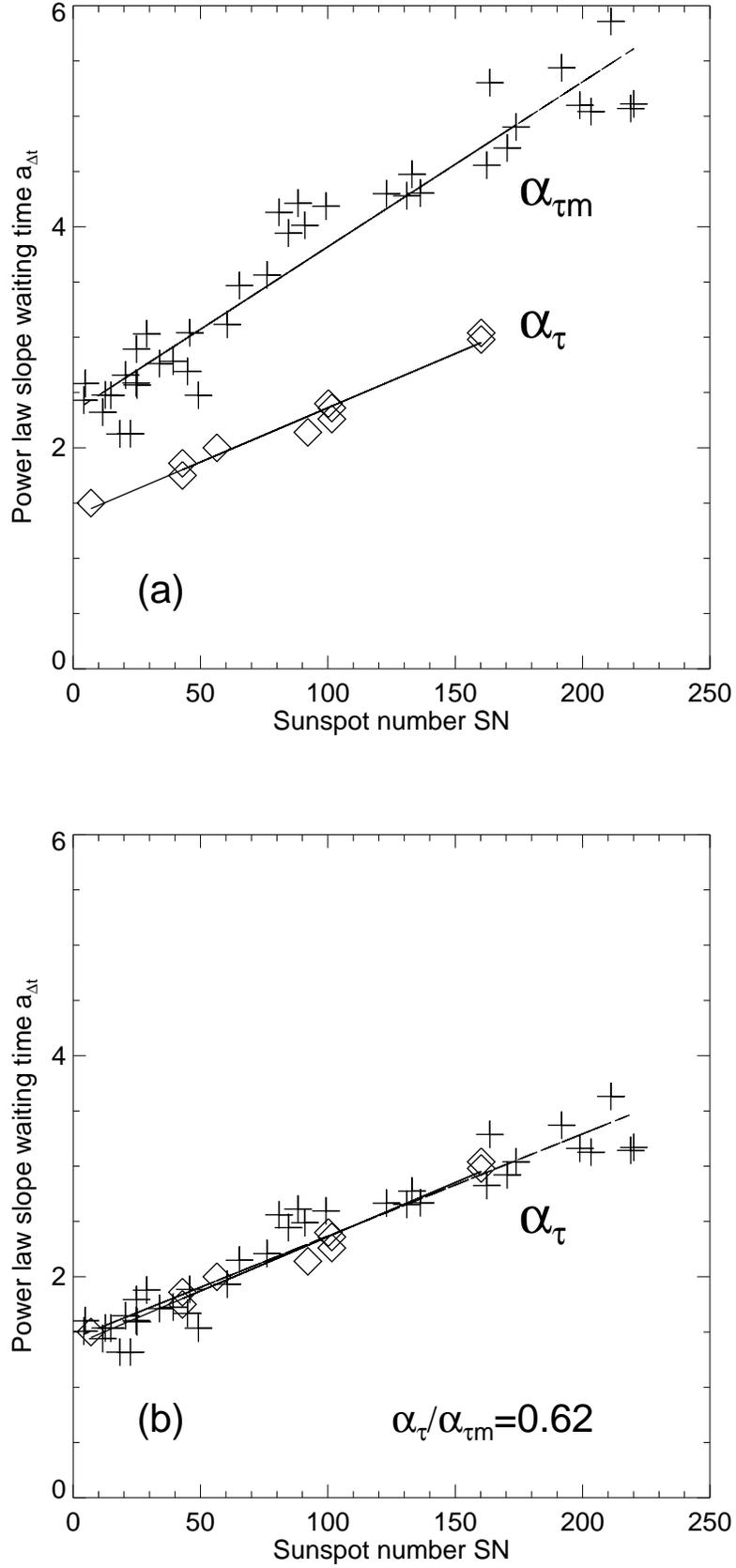}}
\caption{The power law slope $\alpha_\tau$ of the waiting 
time distribution is obtained with standard power law 
fitting methods (diamond symbols), as well as with a 
Pareto-fitting method, $\alpha_{\tau m}$ (cross symbols). 
The Pareto bias correction amounts to an
empirical factor of $\alpha_\tau/\alpha_{\tau m}=0.62$.}
\end{figure}

\begin{figure}
\centerline{\includegraphics[width=1.0\textwidth]{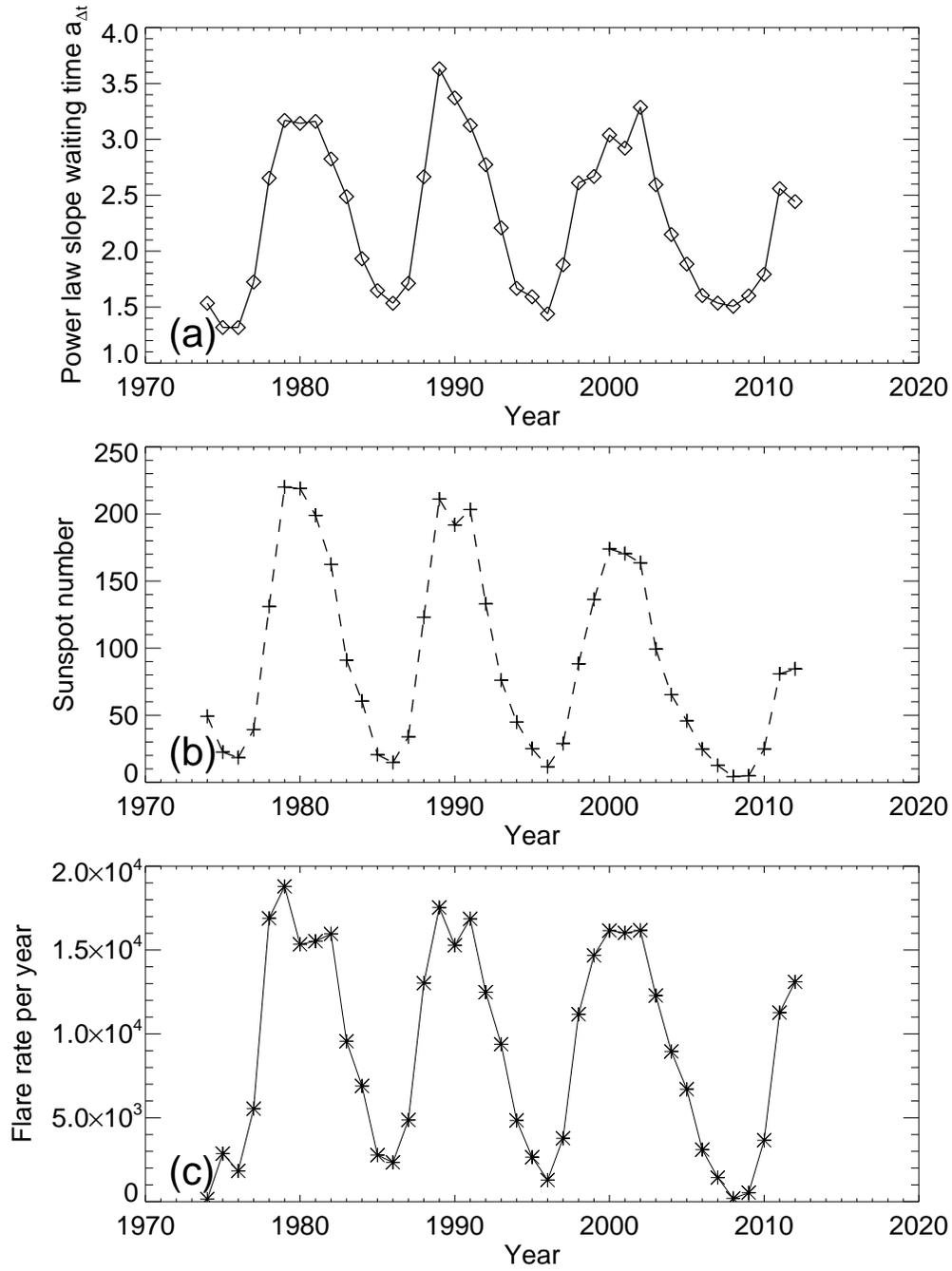}}
\caption{Time evolution of power law slope $\alpha_{\tau}(t)$
of waiting time distributions as a function of the time (a),
time evolution of annual sunspot number during the last
four solar cycles (b), and annual flaring rate $N_{ev}$.}
\end{figure}

\begin{figure}
\centerline{\includegraphics[width=1.0\textwidth]{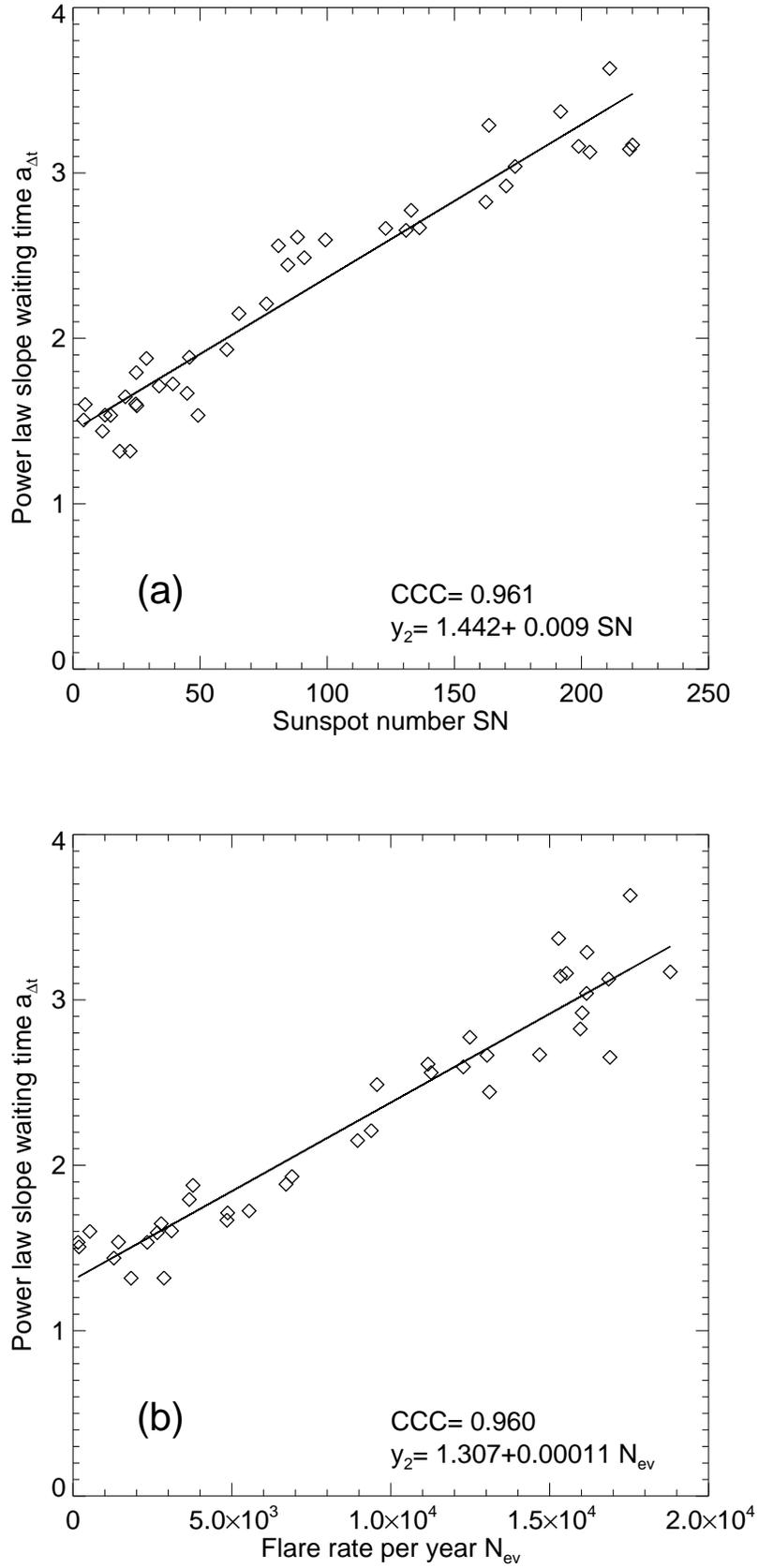}}
\caption{Linear regression fits of the waiting time power 
law slope $\alpha_{\tau}$ versus the sunspot number SN (a)
and versus the annual flare rate $N_{ev}$ (b) 
is shown (solid line), measured from automated flare detections
of GOES flares in annual time intervals (diamonds).} 
\end{figure}

\begin{figure}
\centerline{\includegraphics[width=0.8\textwidth]{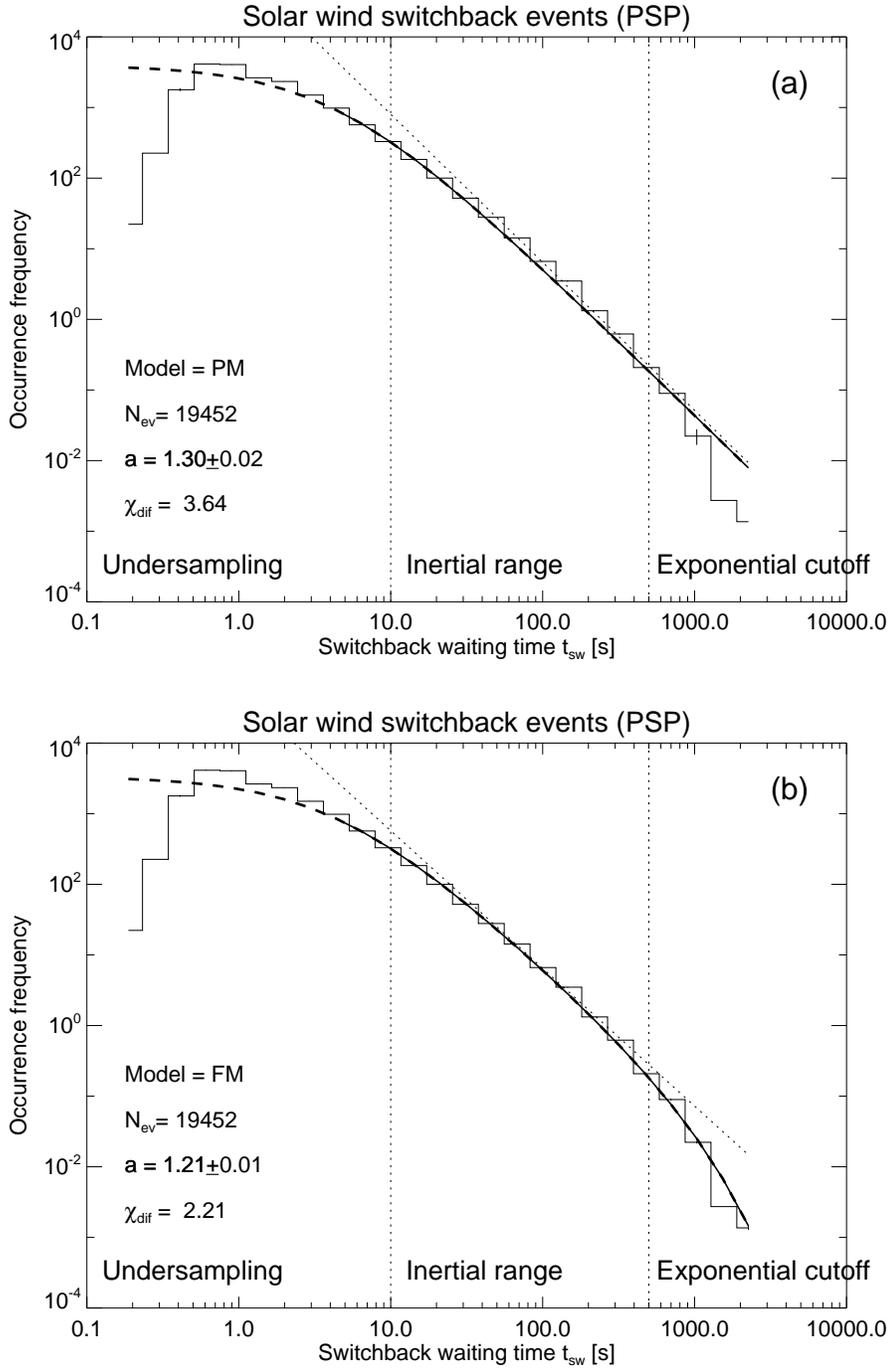}}
\caption{The waiting time distributions of 19,452 magnetic field
switchback events (histograms) observed with the Parker Solar 
Probe. The observed distributions are fitted with two theoretical
models: (a) the Pareto distribution model (PM), and (b) the 
Pareto distribution with an exponential cutoff 
(PF model). Note that the best fit favors the PF model (b) with 
a power law slope of $\alpha_\tau=1.21\pm0.01$ and a 
goodness-of-fit $\chi=2.21$.}
\end{figure}

\begin{figure}
\centerline{\includegraphics[width=1.0\textwidth]{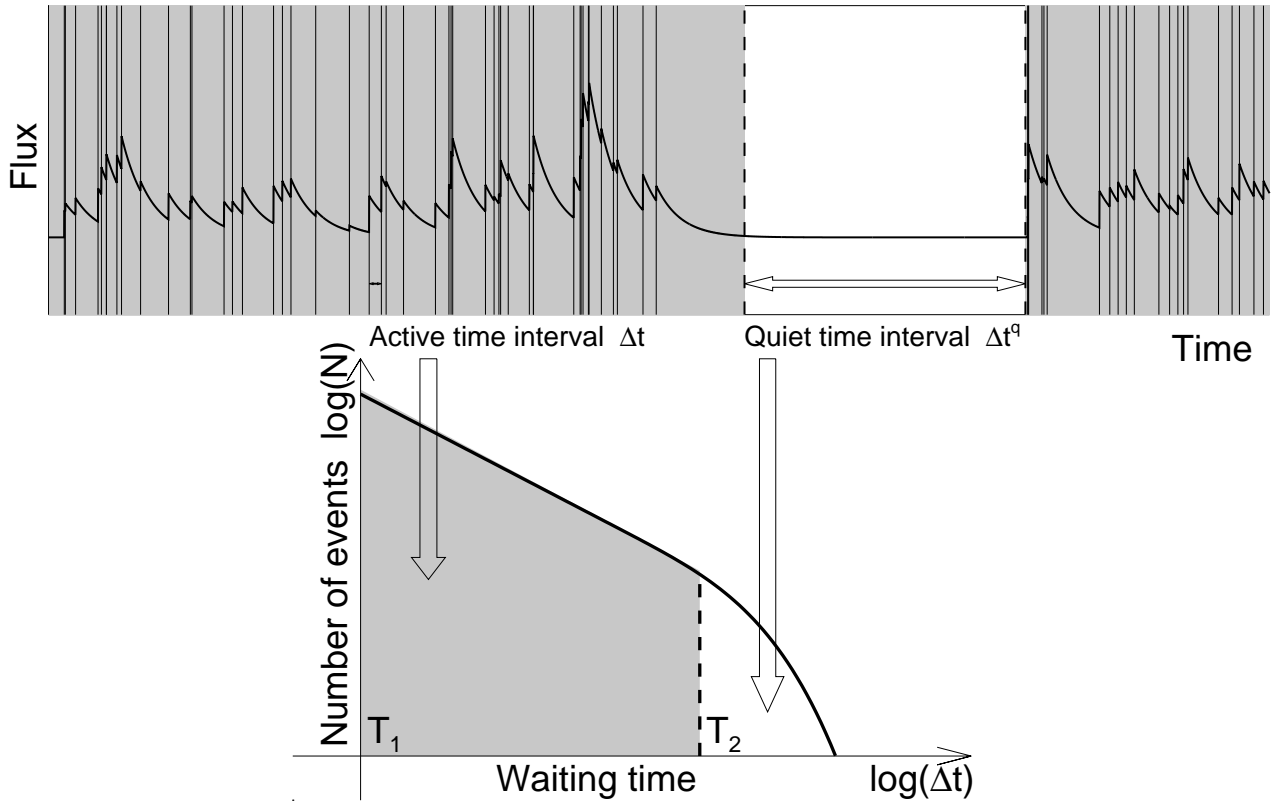}}
\caption{The concept of a dual waiting time distribution is
illustrated, consisting of active time intervals $\Delta t < T$
that contribute to a powerlaw distribution, which is equal 
to that of time duration distributions, $N(T)$. Random-like
quiescent time intervals $\Delta t$ contribute to an
exponential cutoff function. Verticular lines in the upper
panel indicate the start times of events, between which the
waiting times $\Delta t$ are measured (Aschwanden 2014).}
\end{figure}

\begin{figure}
\centerline{\includegraphics[width=0.8\textwidth]{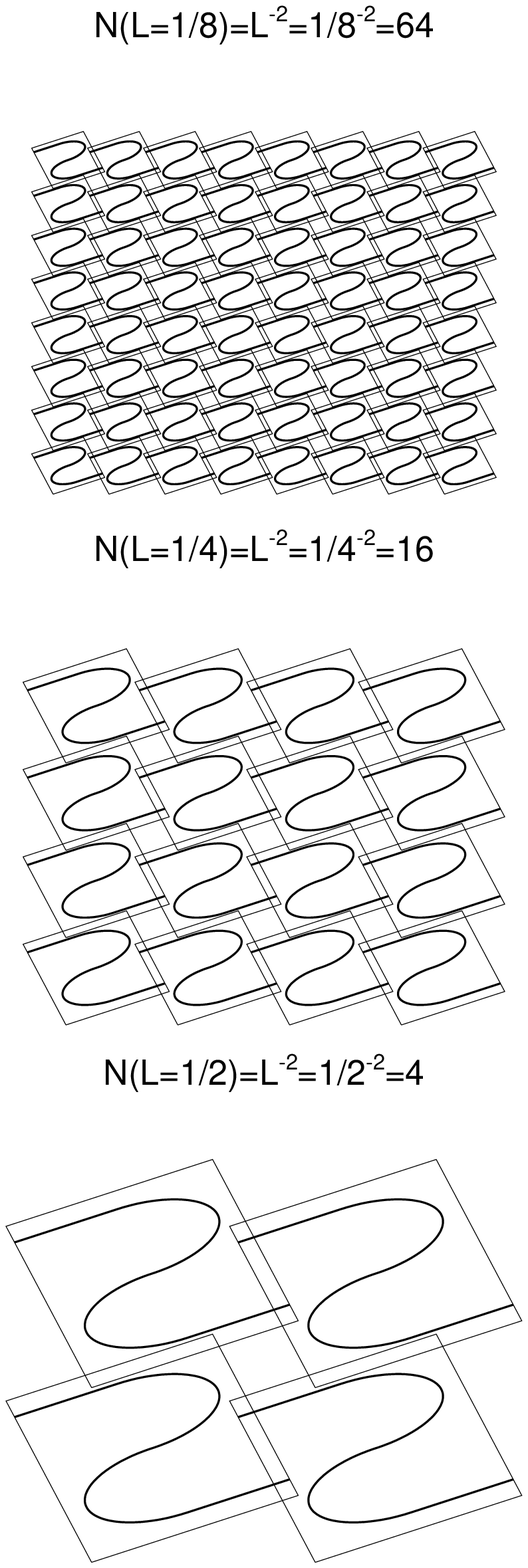}}
\caption{The reciprocal relationship between the geometric
length scales L in two-dimensional Euclidean space 
and occurrence frequency $N(L) \propto L^{-2}$ is depicted 
for three different length scales 
$L=1/8, 1/4, 1/2$, leading to occurrence frequencies of 
$N(L) \propto L^{-2} \propto$ 64 (a), 16 (b), and 4 (c).}
\end{figure}
 
\end{document}